\documentclass{article}
\usepackage{spconf,amsmath,amssymb,graphicx}
\usepackage{booktabs}          % REQUIRED: \toprule/\midrule/\bottomrule
\usepackage{makecell}          % two-line table headers
\usepackage{xcolor}            % red placeholders while drafting
\usepackage[hidelinks]{hyperref}   % load last
\usepackage{enumitem}

\title{From Detection to Anticipation: Continuous Onset Forecasting for Stuttered Speech}
\address{Author Affiliation(s)}
\name{Tampu Ravi Kumar$^{1}$, Ayoub Daliri$^{2}$, Vivek Gupta$^{1}$}
\address{$^{1}$School of Computing and Augmented Intelligence, Arizona State University, USA\\
         $^{2}$College of Health Solutions, Arizona State University, USA\\
         \small{\texttt{\{traviku2, ayoub.daliri, vgupta140\}@asu.edu}}}
\begin{document}
\ninept
\maketitle

\begin{abstract}
Automatic stuttering systems primarily detect dysfluencies after they have occurred, limiting their use for anticipatory intervention. We address this gap by formulating stuttering prediction as \emph{continuous causal time-to-event forecasting}, where a model estimates the risk of an upcoming stuttering onset every 100 ms using only preceding speech. To support this task, we augment  FluencyBank Timestamped for 36 speakers with manually reviewed repetitions, prolongations, and blocks, yielding 2,872 forecasting events. We then model the temporal evolution of frozen Whisper-large-v3 speech representations using a compact causal Conformer with 3.67M trainable parameters. The proposed approach achieves an AUROC of 0.63. Temporal analysis further shows that explicitly modeling preceding speech states outperforms simply increasing the static acoustic window, while a discrete-time survival formulation provides coherent forecasts from 300 ms to 1.5 s using a single model. These results provide initial evidence that stuttering onset is preceded by measurable temporal information in speech and establish a framework for future anticipatory stuttering systems.

\end{abstract}


% ============================================================
\section{Introduction}
\label{sec:introduction}

Developmental stuttering is a speech fluency disorder characterized by repetitions, prolongations, and blocks, affecting approximately 1\% of adults and 5-8\% of children~\cite{yairi2013epidemiology}. Behavioral therapy and altered auditory feedback (AAF) can improve fluency, but treatment responses vary substantially across individuals and situations. Wearable AAF devices such as SpeechEasy provide continuous modified auditory feedback during speech and show heterogeneous benefits during long-term, naturalistic use~\cite{pollard2009speecheasy,ritto2016speecheasy}. A system capable of anticipating stuttering onset could eventually trigger such feedback only when needed rather than continuously.

Most computational work, however, focuses on \emph{stuttering detection}: models classify speech segments that already contain a dysfluency~\cite{lea2021sep28k,bayerl2023dysfluencies,shih2024word}. Recent work extends this setting by predicting whether the next three-second speech segment will contain stuttering~\cite{kozak2026predicting}, but segment-level prediction neither localizes event onset nor continuously estimates its imminence. Motivated by behavioral and neurophysiological evidence that information associated with stuttering can precede overt onset~\cite{jackson2015responses,sowman2012reduced,daliri2018stuttering}, we ask: \emph{can stuttering onset be forecast continuously from speech observed strictly before the event?}

To our knowledge, we are the first to formulate stuttering prediction as \emph{continuous causal time-to-event forecasting}. Every 100 ms, the model observes only preceding speech and estimates the probability that a new stuttering event will begin within a future horizon. We study this problem using FluencyBank Timestamped(FBT)~\cite{romana2024fluencybank} dataset, augmented with manually reviewed repetitions, prolongations, and blocks because the existing annotations are not comprehensive for our forecasting analysis. We represent local speech using frozen Whisper-large-v3~\cite{radford2023whisper} and model its temporal evolution using causal architectures. Our contributions are:
\begin{itemize}[leftmargin=*, nosep]
\item We formulate stuttering anticipation as \textbf{continuous causal
time-to-event forecasting}, producing onset-risk estimates every 100\,ms from
speech observed strictly before the event.
\item We build the \textbf{onset-aligned annotation layer} on existing FBT annotations: 2{,}872 forecasting events across 36 FBT
speakers, including 129 prolongations and 150 blocks annotated manually.
\item We characterize forecastability across event types, speakers, and
horizons from 300\,ms to 1.5\,s, and evaluate forecasts at the \textbf{event
level under an operational duty-cycle constraint}, covering 58.6\% of events
with feedback active only 25\% of the time.
\end{itemize}

\section{TASK FORMULATION AND DATA}
\label{sec:task}

\subsection{Data and Annotation Augmentation}

We use the FBT dataset, comprising
36 speakers with 6.41\,h of recordings and 5.13\,h of transcribed participant
speech. FBT provides word-level timestamps and repetition annotations, but its
labels are not comprehensive for the event types our forecasting task
requires, so we augment it with manually reviewed repetitions, prolongations,
and blocks. We first generate verbatim word-level transcripts and timestamps
with CrisperWhisper2.0~\cite{Zusag_Wagner_Thallinger_2024} and align them with the existing FBT annotations. A
human annotator then listens to each recording end-to-end, rather than
reviewing only automatically flagged regions, and adds missed sound and word
repetitions, prolongations, and blocks; automatic transcripts and existing
annotations serve only as review aids, and all added events are manually
verified. To estimate reliability, a second annotator independently labeled 185 clips
from four speakers (44 repetition,
60 prolongation, 56 block, and 25 event-free clips), without access to the
first annotator's labels. Following SEP-28k~\cite{lea2021sep28k}, we compute
Fleiss' $\kappa$ per event type on clip-level labels: $\kappa = 0.54$
[0.37, 0.69] for repetitions, $0.73$ [0.62, 0.83] for prolongations, and
$0.56$ [0.41, 0.69] for blocks. For matched events, the median absolute
onset discrepancy was 80\,ms overall.
\begin{figure}[t]
\centering
\includegraphics[width=\linewidth]{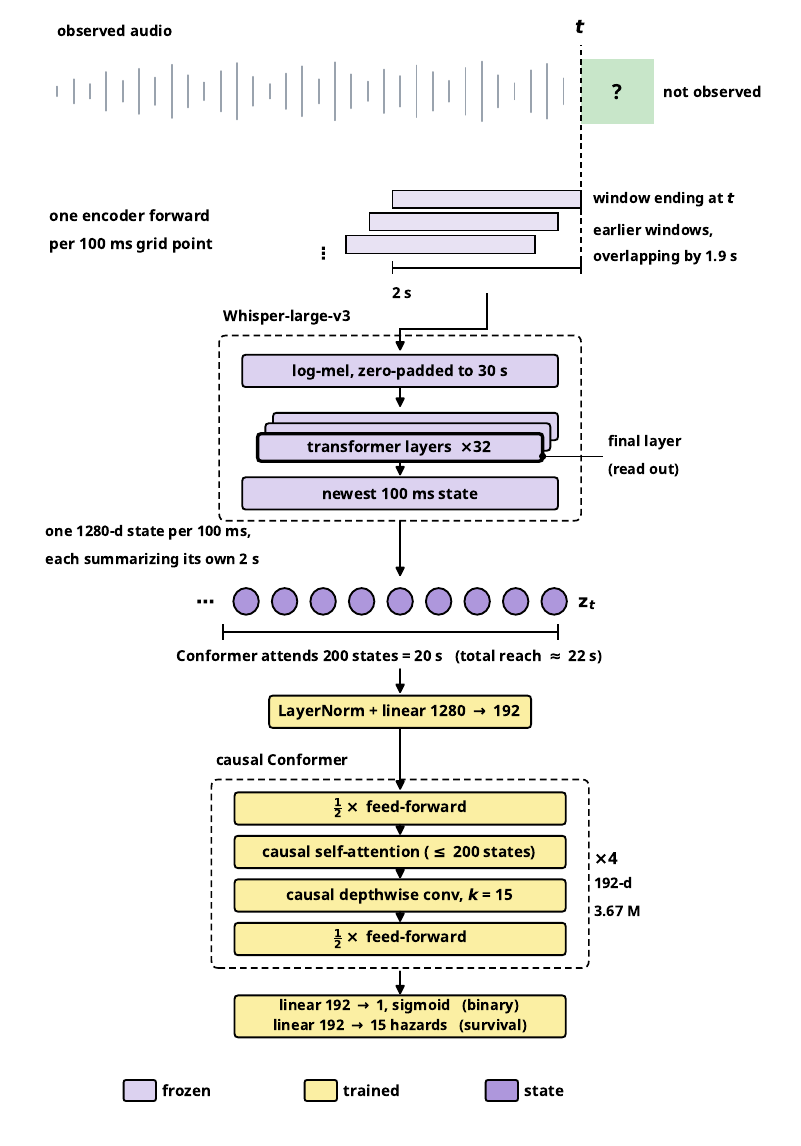}
\caption{Continuous onset forecasting. The encoder is run once per $100$-ms
grid point over its own preceding $2$\,s, so consecutive windows overlap by
$1.9$\,s; the causal Conformer then attends over $200$ such states ($20$\,s),
giving a total reach of $\approx 22$\,s. The $2$\,s window sets how finely
audio is summarized, not how far back the model can see. Nothing the model
reads extends beyond $t$.}

\label{fig:architecture}
\end{figure}

We place event boundaries at the acoustic manifestation of the dysfluency:
repetition onset at the beginning of the first repeated fragment,
prolongation onset at the beginning of the abnormally sustained sound, and
block onset at the point where speech stops or an articulatory posture is
established, i.e.\ the start of the silent or tense interruption rather than
the onset of the subsequent word. Temporally adjacent dysfluencies that
overlap, or that occur within 0.5\, s without an intervening completed fluent
word, are consolidated into a single forecasting event, merging 73 of 2{,}945 raw
dysfluency annotations. The resulting inventory contains 2{,}872 events: 2{,}593
repetitions, 129 prolongations, and 150 blocks, the latter two annotated
entirely by hand. Per-speaker event counts range from 11 to 210 (median 65).
CrisperWhisper is not used by the acoustic forecasting model.

\subsection{Forecasting Target}

Conventional stuttering detection determines whether an observed speech
segment already contains a dysfluency. We instead predict whether a new event will \emph{begin} in the future. 

For each recording, we construct a regular prediction grid at 100-ms
intervals. The grid begins at $t = 2\,$s. The earliest annotated onset in the corpus occurs at 3.37\,s, so no event is lost to the grid start. Let $s^{+}(t)$ denote the onset of the next
repetition, prolongation, or block beginning strictly after prediction time
$t$. We define the time to the next event as
\begin{equation}
  T_t = s^{+}(t) - t.
  \label{eq:tte}
\end{equation}
For a future horizon $\Delta$, the forecasting target is
\begin{equation}
  y_t^{(\Delta)} = \mathbf{1}\!\left[0 < T_t \le \Delta\right].
  \label{eq:target}
\end{equation}
The model estimates
\begin{equation}
  p_t^{(\Delta)} = P\!\left(T_t \le \Delta \mid \mathbf{x}_{\le t}\right),
  \label{eq:risk}
\end{equation}
using only speech observed at or before $t$. Thus, the target event itself is
never contained in a positive input.

% \textbf{Scorable frames.} Of 229{,}966 grid points, we exclude 20{,}107 frames
% falling inside an active stuttering event, because the event has already begun
% and the task would become detection rather than forecasting, and \tbd{} frames
% outside transcribed participant-speech segments. Silence occurring
% \emph{within} a transcribed segment remains scorable, since a deployed system
% would have to produce a decision there. This yields 164{,}694 scorable frames.
% At the primary $\Delta = 300\,$ms horizon, 8{,}121 frames are positive, a
% prevalence of 4.93\%.

\textbf{Scorable frames.} Of 229{,}966 grid points, we exclude 20{,}107 frames falling inside an active stuttering event, because the event has already begun and the task would become detection rather than forecasting, and a further 45{,}165 frames outside transcribed participant-speech segments. Silence occurring \emph{within} a transcribed segment remains scorable, since a deployed system would have to produce a decision there. This yields 164{,}694 scorable frames. At the primary $\Delta = 300$ms horizon, 8{,}121 frames are positive, a prevalence of 4.93\%.

\textbf{Composition of the negative class.} Because stuttering events cluster in time, a model could appear to forecast by recognizing that an event occurred recently rather than by extracting pre-onset information. 25.4\% of positive and 19.8\% of negative frames fall within 2\,s of a previous onset, so the two classes are comparably situated in this respect.

\subsection{Multi-Horizon Formulation}

A fixed binary horizon imposes a hard boundary on a potentially evolving pre-onset state. We therefore additionally evaluate forecasting over $\Delta\in{(0.3,0.5,0.75,1.0,1.5)}$ s. A discrete-time survival model predicts 15 conditional hazards at 100-ms intervals from a single forward pass. Let
$$h_{t,k} = P(T_t \in (\tau_{k-1},\tau_k] \mid T_t > \tau_{k-1}, \mathbf{x}_{\le t}),$$
the probability of onset in bin $k$ given that no onset has occurred earlier, where $\tau_k = 0.1k$s. The cumulative probability of onset within $\tau_k$ is
$$P(T_t \le \tau_k \mid \mathbf{x}_{\le t}) = 1 - \prod_{j=1}^{k}(1-h_{t,j}).$$
The model is trained by masked binary cross-entropy over at-risk bins, i.e.\ the negative log discrete-time survival likelihood. A frame is at risk in bin $k$ only if that bin lies within the transcribed segment containing $t$; otherwise it is right-censored and contributes no further bins, rather than being silently labelled negative as the fixed-horizon target requires. The evaluable set therefore shrinks with the horizon (156{,}669 frames at 0.3\,s to 128{,}167 at 1.5\,s).

This formulation produces a coherent, non-decreasing risk trajectory from a single model rather than requiring a separate binary classifier for every horizon.

% ============================================================
\section{Causal Temporal Forecasting}
\label{sec:model}

% We separate local acoustic representation from longer-range temporal
% modeling (Fig.~\ref{fig:architecture}): at every 100-ms point a frozen
% encoder represents only audio observed before that point, and a lightweight
% causal model integrates the resulting sequence of speech states.

We separate local acoustic representation from longer-range temporal
modeling (Fig.~\ref{fig:architecture}). Every 100\,ms, a frozen encoder turns
the preceding 2\,s of audio into a single vector; a lightweight causal model
then runs over the sequence of those vectors, attending back up to 200 steps.
The 2\,s window therefore sets how finely the audio is summarized, not how far
back the model can see: at a 200-step span the receptive field reaches roughly
22\,s, as 2\,s summaries spaced 100\,ms apart. Everything the model reads at
time $t$ ends at or before $t$.

% \subsection{Acoustic Representation}

% We use the frozen encoder of Whisper-large-v3~\cite{radford2023whisper}.
% Because it applies bidirectional self-attention, encoding a full recording
% would expose future audio, so we independently encode only the 2-s waveform
% ending at $t$, giving $\mathbf{z}_t \in \mathbb{R}^{1280}$. The encoder
% expects a 30-s log-mel input; we zero-pad and retain only the 100 frames
% corresponding to real audio. Frames are emitted at 50\,Hz and averaged in
% non-overlapping groups of five into one state per 100\,ms, matching the
% forecasting grid, giving the causal sequence $\mathbf{Z}_{\le t}$. We use the
% final encoder layer,, \tbd{}[selected on validation /
% without a layer sweep]; layer-wise probes of stuttering \emph{detection}
% locate dysfluency information in middle
% layers~\cite{shih2024word,batra2025layerwise}, but forecasting an event
% absent from the input is a different task. Representations are extracted once
% and cached (\tbd{} GPU-hours), after which each temporal model trains in
% minutes on one GPU.

% ---

\subsection{Acoustic Representation}

We use the frozen encoder of Whisper-large-v3.
Because it applies bidirectional self-attention, encoding a full recording
would expose future audio, so we independently encode only the 2-s waveform
ending at $t$, giving $\mathbf{z}_t \in \mathbb{R}^{1280}$. The encoder
expects a 30-s log-mel input; we zero-pad and retain only the 100 frames
corresponding to real audio. Frames are emitted at 50\,Hz and averaged in
non-overlapping groups of five into one state per 100\,ms, matching the
forecasting grid, giving the causal sequence $\mathbf{Z}_{\le t}$. We use the
final encoder layer, selected on validation average precision from
$\{8, 16, 24, 32\}$ in a preliminary sweep; layer-wise probes of stuttering
\emph{detection} locate dysfluency information in middle
layers~\cite{shih2024word,batra2025layerwise}, but forecasting an event absent
from the input is a different task, and here the final layer was best.
Representations are extracted once and cached (0.9 GPU-hours), after which
each temporal model trains in minutes on one GPU.

% \subsection{Causal Temporal Model}

% Our primary temporal model is a four-block
% Conformer~\cite{gulati2020conformer}. Each Whisper state is normalized and
% projected to 192 dimensions; each block has macaron feed-forward modules
% ($192 \to 768 \to 192$), four-head causal self-attention with relative
% position encoding, and a causal depthwise convolution with kernel size 15.
% Three choices prevent future information from reaching a prediction:
% attention sees only current and preceding states, convolutions are
% left-padded, and we use LayerNorm rather than temporal BatchNorm, which would
% normalize using whole-sequence statistics. Attention uses a maximum left
% context of 20\,s (200 states). The final state maps to a frame-level
% onset-risk probability; the model has 3.67\,M trainable parameters and
% Whisper remains frozen. We compare against a static Whisper predictor that
% mean-pools the 2-s window, a causal TCN (\tbd\,M parameters), and a causal
% GRU (\tbd\,M), all sharing representations, targets, and folds, and we widen
% the static window from 2.0 to 3.8\,s to separate temporal modeling from
% simply observing more audio.

\subsection{Temporal Model}

Our primary temporal model is a four-block
Conformer~\cite{gulati2020conformer}. Each Whisper state is normalized and
projected to 192 dimensions; each block has macaron feed-forward modules
($192 \to 768 \to 192$), four-head causal self-attention, and a causal
depthwise convolution with kernel size 15. Three choices prevent future
information from reaching a prediction: attention sees only current and
preceding states, convolutions are left-padded, and we use LayerNorm rather
than temporal BatchNorm, which would normalize using whole-sequence
statistics. Attention uses a maximum left context of 20\,s (200 states) and
carries no positional term, so ordering reaches the model only through the
stacked causal convolutions. The final state maps to a frame-level onset-risk
probability; the model has 3.67\,M trainable parameters and Whisper remains
frozen. We compare against a static Whisper predictor that attention-pools
the 2-s window, a causal TCN~\cite{Bai2018AnEE} (1.12\,M parameters), and a causal GRU~\cite{cho2014learningphraserepresentationsusing}
(1.12\,M), all sharing representations, targets, and folds. 
% To separate
% temporal modeling from simply observing more audio, we additionally sweep the
% encoder context from 0.5 to 5\,s (Sec.~\ref{sec:context}).

\subsection{Training}

% We optimize focal loss~\cite{lin2017focal} ($\gamma = 2$, $\alpha = 0.75$) to
% counter the 4.93\% prevalence, with AdamW, learning rate $2\times10^{-4}$,
% weight decay $0.01$, dropout $0.2$, gradient clipping at 5, and early
% stopping on validation AP with patience 8. Training sequences span 40\,s: a
% 200-step causal buffer, which supplies context but no loss, followed by 200
% scored steps. For multi-horizon forecasting we keep the architecture and
% replace the binary output with the 15-bin hazard head of
% Eq.~\ref{eq:hazard}.

We optimize focal loss~\cite{lin2017focal} ($\gamma = 2$, $\alpha = 0.75$) to
counter the 4.93\% prevalence, with AdamW~\cite{loshchilov2019}, learning rate $2\times10^{-4}$,
weight decay $0.01$, dropout $0.2$, gradient clipping at 5, and early
stopping on validation AP with patience 8. Training sequences span 40\,s: a
200-step causal buffer, which supplies context but no loss, followed by 200
scored steps. For multi-horizon forecasting we keep the architecture and
replace the binary output with the 15-bin hazard head, trained with plain
binary cross-entropy over at-risk bins rather than focal loss: focal
reweighting distorts predicted probabilities, and calibrated hazards are the
point of that formulation (dropout $0.3$, 50-step buffer).

% ============================================================
\section{Experiments and Results}
\label{sec:results}

\subsection{Setup}

We use six speaker-disjoint folds over the 36 speakers; for each test fold a
disjoint fold is used for validation and the remaining speakers for training.
All models share folds, grids, and targets, so comparisons are paired. We
train three seeds per fold and average their predictions before computing
out-of-fold metrics. Since only 4.93\% of frames are positive, accuracy is
uninformative: we report average precision (AP), its lift over the prevalence
floor, AUROC, and \emph{within-speaker} AUROC (wAUROC), computed within each
speaker and averaged, which removes differences in baseline stuttering
propensity. Intervals and model differences come from paired speaker-level
bootstrap resampling. Operating thresholds are selected on each validation fold and applied unchanged to its test fold. The static Whisper predictor approximates the segment-level setting of \cite{kozak2026predicting} at our temporal resolution.

\subsection{Does Temporal Modeling Improve Forecasting?}

Table~\ref{tab:main} compares architectures at the 300-ms horizon. The causal
Conformer is strongest, with AP $0.0769$ (1.56$\times$ the prevalence floor)
and AUROC $0.6300$, improving pooled AUROC over the TCN by $0.0411$ (95\% CI
$[0.0196, 0.0615]$); the GRU does not improve over the TCN. The gain is not
explained by exposure to more audio: widening the static Whisper window from
2.0 to 3.8\,s moves AUROC from 0.5682 to 0.5666 (AP 0.0522 to 0.0512),
whereas modeling the temporal sequence raises it to 0.6300. Useful pre-onset
information therefore lies in the temporal evolution of locally
contextualized speech states rather than in a larger static window.

\begin{table}[t]
\centering
\caption{Continuous forecasting at a 300\,ms horizon. Lift is AP relative to
the 0.0493 positive prevalence.}
\label{tab:main}
\small
\setlength{\tabcolsep}{4pt}
\begin{tabular}{@{}lcccc@{}}
\toprule
Model & AP $\uparrow$ & Lift $\uparrow$ & AUROC $\uparrow$ & wAUROC $\uparrow$ \\
\midrule
Static Whisper            & 0.0522 & 1.06$\times$ & 0.5682 & 0.5586 \\
GRU                       & 0.0642 & 1.30$\times$ & 0.5587 & 0.5662 \\
TCN                       & 0.0668 & 1.36$\times$ & 0.5889 & 0.5668 \\
\textbf{Causal Conformer} & \textbf{0.0769} & \textbf{1.56$\times$}
                          & \textbf{0.6300} & \textbf{0.5727} \\
\bottomrule
\end{tabular}
\end{table}

\subsection{Event Type, Speakers, and Temporal Context}

Table~\ref{tab:subtype} evaluates each event type against true-negative
frames while excluding frames positive for another type. The Conformer
improves pooled discrimination over the TCN for all three types, reaching
AUROC 0.630, 0.655, and 0.612 for repetitions, prolongations, and blocks.

\begin{table}[t]
\centering
\caption{AUROC by future event type at the 300\,ms horizon
(2{,}593 repetitions, 129 prolongations, 150 blocks).}
\label{tab:subtype}
\small
\setlength{\tabcolsep}{5pt}
\begin{tabular}{@{}lccc@{}}
\toprule
Model & Repetition & Prolongation & Block \\
\midrule
Static Whisper     & 0.574 & 0.508 & 0.497 \\
TCN                & 0.592 & 0.596 & 0.538 \\
\textbf{Conformer} & \textbf{0.630} & \textbf{0.655} & \textbf{0.612} \\
\bottomrule
\end{tabular}
\end{table}

Within-speaker analysis is more conservative. Repetition forecasting stays
above chance, with mean wAUROC $0.576$ (95\% CI $[0.564, 0.590]$) across all
36 speakers and 97\% of speakers above 0.5. For prolongations and blocks mean
wAUROC is 0.540 and 0.541, with intervals overlapping chance and sufficient
positive frames in only 17 and 23 speakers. The gap between pooled and
within-speaker discrimination indicates substantial speaker heterogeneity,
particularly for the rarer types.

Varying the causal attention history while holding the 2-s Whisper
representation fixed, a 2-s history yields AUROC 0.592 and a 5-s history
0.621, with no measurable gain beyond (0.620 at 10\,s, 0.622 at 20\,s).
Useful temporal information therefore extends past the local acoustic window
but saturates near 5\,s.

\subsection{Multi-Horizon and Operational Forecasting}

Table~\ref{tab:horizon} compares the discrete-time hazard model with
independently trained binary classifiers and with independent cumulative
outputs. The hazard model is competitive with five separately trained binary
models while producing all horizons from one predictor. Independent
cumulative outputs reach similar discrimination but violate the required
ordering $P(T \le \Delta_i) \le P(T \le \Delta_j)$ for $\Delta_i < \Delta_j$
on 68.9\% of frames, whereas the survival formulation is coherent by
construction.

\begin{table}[t]
\centering
\caption{Multi-horizon forecasting (AUROC\,/\,AP). The hazard model produces
all horizons from one trained model and is monotone by construction. This table uses a censoring-aware evaluation set}
\label{tab:horizon}
\small
\setlength{\tabcolsep}{4pt}
\begin{tabular}{@{}lccc@{}}
\toprule
$\Delta$ & Hazard & Binary & \makecell{Independent\\cumulative} \\
\midrule
0.30\,s & .604/.074 & .598/.074 & .610/.075 \\
0.50\,s & .583/.114 & .582/.115 & .593/.118 \\
0.75\,s & .577/.177 & .593/.183 & .591/.183 \\
1.00\,s & .576/.218 & .572/.215 & .589/.224 \\
1.50\,s & .575/.310 & .588/.318 & .585/.317 \\
\bottomrule
\end{tabular}
\end{table}

Frame-level discrimination does not show whether forecasts could drive an
event-triggered intervention. We therefore merge positive decisions into
warning episodes and measure event-level sensitivity against alarm duty
cycle, the fraction of time a feedback system would be active. At a 25\% duty
cycle the Conformer anticipates 58.6\% of scorable events, against 57.4\% for
the TCN and 53.8\% for the GRU; at a higher-recall point it anticipates 91.6\%
at 66.5\% duty. At the validation-selected maximum-F1 threshold it anticipates
43.9\% of events with a mean lead time of 196\,ms, but produces 27.6 false
warning bouts per minute. These operating points quantify the trade-off
between feedback exposure and event coverage; whether selectively triggered
AAF at these lead times improves fluency requires prospective intervention
studies.

% ============================================================

\section{Related Work}
\label{sec:related}

Forecasting is plausible because stuttering is not distributed uniformly
through speech: it concentrates on particular sounds, word classes, and
utterance positions~\cite{brown1945loci,warner2023linguistic},
and speakers often report anticipating upcoming moments of
stuttering~\cite{jackson2015responses}. Neural
differences precede overt blocks~\cite{sowman2012reduced}, and adults who
stutter show reduced pre-speech auditory modulation~\cite{daliri2018stuttering}.
Das et al.~\cite{das2020facial} predicted upcoming dysfluent trials from
pre-speech facial muscle movements. The listener study of Liu and
Howell~\cite{liu2024listeners} and the surprisal analysis of Warner et
al.~\cite{warner2026information} are closest to our setting and both ask whether pre-onset material predicts the event, but
neither builds an acoustic predictor.

The only audio-based predictor we are aware of is that of
Kozak~\cite{kozak2026predicting}, who trains a compact CNN on SEP-28k to
predict whether the next 3\,s clip contains a dysfluency and finds signal
concentrated on blocks and sound repetitions. Our formulation differs in three
respects: it produces a risk estimate every 100\,ms rather than one decision
per clip, it targets exact annotated onsets at sub-second horizons, and it
evaluates event-level coverage under an alarm duty-cycle budget. More broadly,
continuous causal prediction of upcoming speech events has been studied for
turn-taking, where models project future voice activity from preceding
audio~\cite{skantze2017towards,ekstedt2022vap}; our event-level evaluation,
sensitivity against time spent in warning, follows practice in seizure
prediction~\cite{snyder2008statistics}, and our multi-horizon head uses a
discrete-time survival formulation~\cite{Gensheimer_Narasimhan_2019}.

\section{Conclusion}
\label{sec:conclusion}

In this work, we formulated stuttering anticipation as continuous causal time-to-event forecasting, constructed the required onset-aligned corpus layer, and established a benchmark framework for evaluating pre-onset risk trajectories. Our findings demonstrate three key insights: (1) speech preceding stuttering onset carries statistically measurable predictive structure; (2) modeling the sequential evolution of contextualized speech states provides significant gains over expanding static acoustic context; and (3) a discrete-time survival formulation produces coherent, multi-horizon forecasts without violating temporal ordering. Although current operational trade-offs indicate that pre-onset forecasting remains an open challenge, this work provides a formulation, empirical baseline, and evaluation methodology to guide future research toward deployable anticipatory interventions.

\section{Limitations and Future Work}
\label{sec:limitations}

While this work establishes a baseline for continuous onset forecasting, several challenges remain before such models can drive real-time clinical interventions. The primary constraint remains annotation scale. Continuous time-to-event forecasting requires precise, frame-level onset timestamps that existing segment-level stuttering corpora do not provide. While our augmented layer of 2,872 events across 36 speakers represents a critical first step, scaling onset-aligned annotations to larger, more diverse cohorts will be essential to fully unlock fine-grained anticipatory modeling.

\section{Compliance with Ethical Standards}
This work uses the publicly available FluencyBank Timestamped (FBT) dataset under its existing license and our augmented annotations contain no additional PII beyond what FBT already provides. The proposed forecasting models are intended as a research step toward clinician- or user-supervised assistive tools, not as a standalone diagnostic or autonomous intervention system.

\section{Acknowledgments}
\label{sec:acknowledgments}

We conducted this research as a collaborative effort at Arizona State University. We thank the Complex Data Analysis and Reasoning Lab within the School of Computing and Augmented Intelligence at Arizona State University for providing computational resources and institutional support. We also sincerely thank Manan Roy Choudhury and Tejas Anvekar for their valuable feedback during the ideation stage of this work.

\bibliographystyle{IEEEbib}
\bibliography{strings}

@misc{kozak2026predicting,
      title={Predicting Upcoming Stuttering Events from Three-Second Audio: Stratified Evaluation Reveals Severity-Selective Precursors, and the Model Deploys Fully On-Device},
      author={Nazar Kozak},
      year={2026},
      eprint={2604.27279},
      archivePrefix={arXiv},
      primaryClass={cs.SD},
      url={https://arxiv.org/abs/2604.27279},
}

@InProceedings{radford2023whisper,
  title = 	 {Robust Speech Recognition via Large-Scale Weak Supervision},
  author =       {Radford, Alec and Kim, Jong Wook and Xu, Tao and Brockman, Greg and Mcleavey, Christine and Sutskever, Ilya},
  booktitle = 	 {Proceedings of the 40th International Conference on Machine Learning},
  pages = 	 {28492--28518},
  year = 	 {2023},
  publisher =    {PMLR},
  url = 	 {https://proceedings.mlr.press/v202/radford23a.html}
}

@article{batra2025layerwise,
  title={Exploring Whisper Embeddings for Stutter Detection: A Layer-Wise Study},
  author={Ashita Batra and Brajesh Kar and Pradip K. Das},
  journal={2025 33rd European Signal Processing Conference (EUSIPCO)},
  year={2025},
  pages={61--65},
  url={https://api.semanticscholar.org/CorpusID:283099140}

}

@article{Bai2018AnEE,
  title={An Empirical Evaluation of Generic Convolutional and Recurrent Networks for Sequence Modeling},
  author={Shaojie Bai and J. Zico Kolter and Vladlen Koltun},
  journal={ArXiv},
  year={2018},
  volume={abs/1803.01271},
  url={https://api.semanticscholar.org/CorpusID:4747877}
}

@inproceedings{bayerl2023dysfluencies, title={Detecting Dysfluencies in Stuttering Therapy Using wav2vec 2.0}, url={https://www.isca-archive.org/interspeech_2022/bayerl22b_interspeech.html}, DOI={10.21437/Interspeech.2022-10908}, booktitle={Interspeech 2022}, author={Bayerl, Sebastian Peter and Wagner, Dominik and Noeth, Elmar and Riedhammer, Korbinian}, year={2022}, pages={2868--2872}, language={en} }

@article{daliri2018stuttering, title={Stuttering adults’ lack of pre-speech auditory modulation normalizes when speaking with delayed auditory feedback}, volume={99}, ISSN={00109452}, url={https://linkinghub.elsevier.com/retrieve/pii/S0010945217303672}, DOI={10.1016/j.cortex.2017.10.019}, journal={Cortex}, author={Daliri, Ayoub and Max, Ludo}, year={2018}, month=feb, pages={55--68}, language={en} }

@article{Gensheimer_Narasimhan_2019, title={A scalable discrete-time survival model for neural networks}, volume={7}, rights={http://creativecommons.org/licenses/by/4.0/}, ISSN={2167-8359}, url={https://peerj.com/articles/6257}, DOI={10.7717/peerj.6257}, abstractNote={There is currently great interest in applying neural networks to prediction tasks in medicine. It is important for predictive models to be able to use survival data, where each patient has a known follow-up time and event/censoring indicator. This avoids information loss when training the model and enables generation of predicted survival curves. In this paper, we describe a discrete-time survival model that is designed to be used with neural networks, which we refer to as Nnet-survival. The model is trained with the maximum likelihood method using mini-batch stochastic gradient descent (SGD). The use of SGD enables rapid convergence and application to large datasets that do not fit in memory. The model is flexible, so that the baseline hazard rate and the effect of the input data on hazard probability can vary with follow-up time. It has been implemented in the Keras deep learning framework, and source code for the model and several examples is available online. We demonstrate the performance of the model on both simulated and real data and compare it to existing models Cox-nnet and Deepsurv.}, journal={PeerJ}, author={Gensheimer, Michael F. and Narasimhan, Balasubramanian}, year={2019}, pages={e6257}, language={en} }

@inproceedings{gulati2020conformer, title={Conformer: Convolution-augmented Transformer for Speech Recognition}, url={https://www.isca-archive.org/interspeech_2020/gulati20_interspeech.html}, DOI={10.21437/Interspeech.2020-3015}, booktitle={Interspeech 2020}, author={Gulati, Anmol and Qin, James and Chiu, Chung-Cheng and Parmar, Niki and Zhang, Yu and Yu, Jiahui and Han, Wei and Wang, Shibo and Zhang, Zhengdong and Wu, Yonghui and Pang, Ruoming}, year={2020}, pages={5036--5040}, language={en} }

@article{jackson2015responses, title={Responses of adults who stutter to the anticipation of stuttering}, volume={45}, ISSN={0094730X}, url={https://linkinghub.elsevier.com/retrieve/pii/S0094730X15000534}, DOI={10.1016/j.jfludis.2015.05.002}, journal={Journal of Fluency Disorders}, author={Jackson, Eric S. and Yaruss, J. Scott and Quesal, Robert W. and Terranova, Valerie and Whalen, D.H.}, year={2015}, pages={38--51}, language={en} }

@inproceedings{lea2021sep28k, title={SEP-28k: A Dataset for Stuttering Event Detection from Podcasts with People Who Stutter}, rights={https://ieeexplore.ieee.org/Xplorehelp/downloads/license-information/IEEE.html}, ISBN={9781728176055}, url={https://ieeexplore.ieee.org/document/9413520/}, DOI={10.1109/ICASSP39728.2021.9413520}, booktitle={ICASSP 2021 - 2021 IEEE International Conference on Acoustics, Speech and Signal Processing (ICASSP)}, author={Lea, Colin and Mitra, Vikramjit and Joshi, Aparna and Kajarekar, Sachin and Bigham, Jeffrey P.}, year={2021}, pages={6798--6802} }

@inproceedings{lin2017focal, address={Venice}, title={Focal Loss for Dense Object Detection}, ISBN={9781538610329}, url={http://ieeexplore.ieee.org/document/8237586/}, DOI={10.1109/ICCV.2017.324}, booktitle={2017 IEEE International Conference on Computer Vision (ICCV)}, author={Lin, Tsung-Yi and Goyal, Priya and Girshick, Ross and He, Kaiming and Dollar, Piotr}, year={2017}, month=oct, pages={2999--3007} }

@article{pollard2009speecheasy, title={Effects of the SpeechEasy on Objective and Perceived Aspects of Stuttering: A 6-Month, Phase I Clinical Trial in Naturalistic Environments}, volume={52}, ISSN={1092-4388, 1558-9102}, url={http://pubs.asha.org/doi/10.1044/1092-4388%282008/07-0204%29}, DOI={10.1044/1092-4388(2008/07-0204)}, abstractNote={Purpose
              Effects of the SpeechEasy when used under extraclinical conditions over several months were investigated. Primary purposes were to help establish Phase I level information about the therapeutic utility of the SpeechEasy and to compare those results with previous findings obtained in laboratory and clinical settings.
            
            
              Method
              Eleven adults who stutter participated. A nonrandomized ABA group design was utilized. Speech samples were collected every 2 weeks in extraclinical environments. Qualitative data were collected through weekly written logs and an exit questionnaire.
            
            
              Results
              Group analyses revealed a statistically significant effect of the SpeechEasy immediately postfitting but no treatment effect across 4 months’ time. Individual responses varied greatly with regard to stuttering frequency and subjective impressions. Relatively more stuttering reduction occurred during oral reading than during formulated speech.
            
            
              Conclusions
              Based on this protocol, Phase II trials are not indicated. However, positive individual responses and self-reports suggest some clinical utility for the SpeechEasy. The use of more challenging sampling procedures strengthened external validity and captured more modest altered auditory feedback effects compared with those previously reported in laboratory settings. Device use coincided more so with positive subjective impressions than with measurable fluency improvement, highlighting challenges facing clinicians when implementing principles of evidence-based practice, including client-based preferences.}, number={2}, journal={Journal of Speech, Language, and Hearing Research}, author={Pollard, Ryan and Ellis, John B. and Finan, Don and Ramig, Peter R.}, year={2009}, month=apr, pages={516--533}, language={en} }

@article{ritto2016speecheasy, title={Randomized clinical trial: the use of SpeechEasy® in stuttering treatment}, volume={51}, rights={http://doi.wiley.com/10.1002/tdm_license_1.1}, ISSN={13682822}, url={https://onlinelibrary.wiley.com/doi/10.1111/1460-6984.12237}, DOI={10.1111/1460-6984.12237}, number={6}, journal={International Journal of Language \& Communication Disorders}, author={Ritto, Ana Paula and Juste, Fabiola Staróbole and Stuart, Andrew and Kalinowski, Joseph and De Andrade, Claudia Regina Furquim}, year={2016}, month=nov, pages={769--774}, language={en} }

@inproceedings{shih2024word, title={Self-Supervised Speech Models For Word-Level Stuttered Speech Detection}, rights={https://doi.org/10.15223/policy-029}, ISBN={9798350392258}, url={https://ieeexplore.ieee.org/document/10832220/}, DOI={10.1109/SLT61566.2024.10832220}, booktitle={2024 IEEE Spoken Language Technology Workshop (SLT)}, publisher={IEEE}, author={Shih, Yi-Jen and Gkalitsiou, Zoi and Dimakis, Alexandros G. and Harwath, David}, year={2024}, month=dec, pages={937--944} }

@article{sowman2012reduced, title={Reduced activation of left orbitofrontal cortex precedes blocked vocalization: A magnetoencephalographic study}, volume={37}, rights={https://www.elsevier.com/tdm/userlicense/1.0/}, ISSN={0094730X}, url={https://linkinghub.elsevier.com/retrieve/pii/S0094730X1200040X}, DOI={10.1016/j.jfludis.2012.05.001}, number={4}, journal={Journal of Fluency Disorders}, author={Sowman, Paul F. and Crain, Stephen and Harrison, Elisabeth and Johnson, Blake W.}, year={2012}, month=dec, pages={359--365}, language={en} }

@article{yairi2013epidemiology, title={Epidemiology of stuttering: 21st century advances}, volume={38}, ISSN={0094730X}, url={https://linkinghub.elsevier.com/retrieve/pii/S0094730X12001052}, DOI={10.1016/j.jfludis.2012.11.002}, number={2}, journal={Journal of Fluency Disorders}, author={Yairi, Ehud and Ambrose, Nicoline}, year={2013}, month=june, pages={66--87}, language={en} }

@inproceedings{Zusag_Wagner_Thallinger_2024, title={CrisperWhisper: Accurate Timestamps on Verbatim Speech Transcriptions}, url={https://www.isca-archive.org/interspeech_2024/zusag24_interspeech.html}, DOI={10.21437/Interspeech.2024-731}, booktitle={Interspeech 2024}, author={Zusag, Mario and Wagner, Laurin and Thallinger, Bernhard}, year={2024}, pages={1265--1269}, language={en} }

@article{romana2024fluencybank,
author = {Romana, Amrit and Niu, Minxue and Perez, Matthew and Mower Provost, Emily},
year = {2024},
month = {10},
pages = {4203--4215},
title = {FluencyBank Timestamped: An Updated Data Set for Disfluency Detection and Automatic Intended Speech Recognition},
volume = {67},
journal = {Journal of Speech, Language, and Hearing Research},
doi = {10.1044/2024_JSLHR-24-00070}
}

@misc{cho2014learningphraserepresentationsusing,
      title={Learning Phrase Representations using RNN Encoder-Decoder for Statistical Machine Translation}, 
      author={Kyunghyun Cho and Bart van Merrienboer and Caglar Gulcehre and Dzmitry Bahdanau and Fethi Bougares and Holger Schwenk and Yoshua Bengio},
      year={2014},
      eprint={1406.1078},
      archivePrefix={arXiv},
      primaryClass={cs.CL},
      url={https://arxiv.org/abs/1406.1078}, 
}

@misc{loshchilov2019,
      title={Decoupled Weight Decay Regularization}, 
      author={Ilya Loshchilov and Frank Hutter},
      year={2019},
      eprint={1711.05101},
      archivePrefix={arXiv},
      primaryClass={cs.LG},
      url={https://arxiv.org/abs/1711.05101}, 
}

@article{brown1945loci,
  title   = {The loci of stutterings in the speech sequence},
  author  = {Brown, Spencer F.},
  journal = {Journal of Speech Disorders},
  volume  = {10}, number = {3}, pages = {181--192}, year = {1945}
}

@article{warner2023linguistic,
  title   = {Linguistic features of stuttering during spontaneous speech},
  author  = {Warner, Haley J. and Shroff, Ravi and Zuanazzi, Arianna and Arenas, Richard M. and Jackson, Eric S.},
  journal = {Journal of Fluency Disorders},
  volume  = {78}, pages = {106016}, year = {2023},
  doi     = {10.1016/j.jfludis.2023.106016}
}

@misc{das2020facial,
      title={Stuttering Speech Disfluency Prediction using Explainable Attribution Vectors of Facial Muscle Movements},
      author={Das, Arun and Mock, Jeffrey and Chacon, Henry and Irani, Farzan and Golob, Edward and Najafirad, Peyman},
      year={2020},
      eprint={2010.01231},
      archivePrefix={arXiv},
      url={https://arxiv.org/abs/2010.01231},
}

@inproceedings{skantze2017towards,
  title     = {Towards a General, Continuous Model of Turn-taking in Spoken Dialogue using {LSTM} Recurrent Neural Networks},
  author    = {Skantze, Gabriel},
  booktitle = {Proc. SIGDIAL}, year = {2017}
}

@inproceedings{ekstedt2022vap,
  title     = {Voice Activity Projection: Self-supervised Learning of Turn-taking Events},
  author    = {Ekstedt, Erik and Skantze, Gabriel},
  booktitle = {Proc. Interspeech}, year = {2022}
}

@article{snyder2008statistics,
  title   = {The statistics of a practical seizure warning system},
  author  = {Snyder, David E. and Echauz, Javier and Grimes, David B. and Litt, Brian},
  journal = {Journal of Neural Engineering},
  volume  = {5}, number = {4}, pages = {392--401}, year = {2008}
}

@article{liu2024listeners,
  title   = {Can listeners predict whether or not a stutter follows a stretch of fluent speech?},
  author  = {Liu, Xena and Howell, Peter},
  journal = {Journal of Fluency Disorders},
  year    = {2024}
}

@article{warner2026information,
  title   = {Information Load Predicts Stuttering Events},
  author  = {Warner, Haley J. and Bernstein Ratner, Nan and Jackson, Eric S.},
  journal = {Journal of Speech, Language, and Hearing Research},
  volume  = {69}, number = {4}, pages = {1554--1567}, year = {2026},
  doi     = {10.1044/2026_JSLHR-25-00833}
}

\end{document}